\documentclass[12pt]{iopart}
\usepackage{graphicx} % to include figure files
\begin{document}

\def\la{\langle}
\def\ra{\rangle}
\def\om{\omega}
\def\Om{\Omega}
\def\vep{\varepsilon}
\def\wh{\widehat}
\def\P0{\wh{\cal P}_0}
\def\dt{\delta t}
\newcommand{\beq}{\begin{equation}}
\newcommand{\eeq}{\end{equation}}
\newcommand{\beqa}{\begin{eqnarray}}
\newcommand{\eeqa}{\end{eqnarray}}
\newcommand{\intf}{\int_{-\infty}^\infty}
\newcommand{\into}{\int_0^\infty}
\title[Time-of-arrival measurements]{Atomic time-of-arrival measurements with
a laser of finite beam width}
\author{J. A. Damborenea*\dag, I. L. Egusquiza*,
G. C. Hegerfeldt\ddag\, and J. G. Muga\dag}
\address{* Fisika Teorikoaren Saila, Euskal Herriko Unibertsitatea,
644 P.K., 48080 Bilbao, Spain}
\address{\dag Departamento de Qu\'\i{}mica-F\'\i{}sica, Universidad del
Pa\'\i s Vasco, Apdo. 644, 48080 Bilbao, Spain}
\address{\ddag Institut f\"ur Theoretische Physik, Universit\"at
G\"ottingen, Bunsenstr. 9, 37073 G\"ottingen, Germany}

\begin{abstract}
A natural approach to measure the time of arrival of an atom 
at a spatial region is to illuminate this region with a laser and 
detect the first fluorescence photons produced by the excitation of the 
atom 
and subsequent decay.  
We investigate the actual physical content of such a measurement
in terms of atomic dynamical variables, 
taking into account the finite width of the laser beam.  
%for the case where only the first photon emited 
%by the atom is recorded, or equivalently for atomic level 
%configurations where only one photon is produced per atom.  
Different operation regimes are identified, in particular 
the ones in which the quantum current density may be obtained. 
%For strong driving and in the   
%limit of a large lifetime the quantum current density may be obtained 
%by deconvolution.      
\end{abstract}
\pacs{03.65.-w, 42.50-p, 32.80-t}
\maketitle
\section{Introduction}

In spite of the emphasis on ``observables'' and ``measurements''
in the standard formulation of quantum mechanics, 
one of the fundamental difficulties of the theory is to establish the
relation between 
``observables'' represented by operators that act on the
variables 
of the microscopic system in isolation,   
and  actual measurements.
There is no universal prescription to construct
an apparatus for 
a given operator and, in the other direction, identifying from the 
macroscopic experiment the corresponding 
microscopic operator may also be difficult. These problems have been
particularly accute, or have been perceived to be so, 
for time observables. According to an argument due to
Pauli \cite{Pauli},
there is no
self-adjoint time operator conjugate to a semibounded Hamiltonian
(see however \cite{Galapon}), and this limitation has hindered  
the research on the
theoretical 
understanding and description of time quantities measured in the
laboratories. 
The trend is slowly changing though \cite{MSE02}.  
During the last two decades, many works have 
been devoted to find consistent
theories, in terms of operators or otherwise, about characteristic
times
for tunneling, quantum jumps, decay, or arrival \cite{ML00,MSE02}.
This effort has been mostly focused  on establishing fundamental
theories for the isolated system, typically  
a structureless particle. In these theories the
measuring apparatus
is totally absent, at least in an explicit form, 
or at best modelled very schematically \cite{AOPRU98,Halliwell98b}.
We have 
advocated the need for an analysis of more realistic models
\cite{ML00,DEHM02},
not only to 
determine how and if the ideally defined quantities can be measured
but also to ascertain 
what usual operational methods are ``really'' measuring in the 
language of
system variables. 
%The analysis is also motivated by the current technical
%ability
%to manipulate 
%ultracold atoms, where divergence between classical and quantum results 
%may be expected \cite{ML00}. 
In a 
previous article \cite{DEHM02} we have modelled the measurement by
fluorescence of the time of arrival (TOA) of an atom at a given location.
An important finding was that, in the limit of 
{\it weak driving}, the quantum current density $J$ may be obtained from the
distribution of first photon 
detections $\Pi$. This is satisfactory because, in the classical
limit, $J$ is the correct TOA distribution 
for the initial states considered (with momenta fully directed towards
the laser); but    
it is also surprising because $J$ may be negative in quantum mechanics, 
even for states formed entirely by positive momentum components
\cite{backflow}. By contrast, the positive and axiomatic TOA distribution
of Kijowski \cite{Kijowski}
was not obtained in any limit of the parameters involved,
except when it is sufficiently close to 
$J$.  
A simplifying assumption of the model 
was that the laser-illuminated region 
had a semi-infinite width, from $x=0$ to $\infty$. This is a good 
approximation at very low kinetic energies, because the atom is either detected 
before abandoning the finite illuminated region,
or reflected by the laser;  
but it fails otherwise. 
In this paper we
remove that
assumption and consider the more realistic case of a finite laser beam
width, $L$. 

There were two basic difficulties with a direct ``good TOA measurement''
in the semi-infinite
model: the atomic reflection with no photon emission, due to a strong laser  
field, and the detection delay for a weak one. 
Although we could find and characterize operation regimes where the
two effects
were 
negligible in practice, the best fit with the free-motion current density   
was obtained 
for weak driving and for a short lifetime $1/\gamma$ of the excited state 
($\gamma$ is Einstein's coefficient); this avoids reflection 
but induces delay because, in spite of the short life time,
the laser pumping 
is very inefficient. 
%The key point is that, when the atoms are reflected without being detected,
%the 
%information on reflected low
%energy components is lost. On the other hand,  without     
%reflection, all the relevant information is kept, even if the measured
%signal is 
%distorted by delays. This means that  
Nevertheless, the
flux $J$ for the freely moving atoms may be obtained
in principle from the 
experimental signal $\Pi$, since, in that limit, 
$\Pi=J*W$, 
where $W$ is  
the (known) distribution of 
photon detection times corresponding to a laser-illuminated
atom at rest.  
 
The applicability of the above ``deconvolution'' method 
is limited because the long detection delays implied require 
also a large laser-beam width, and the required width increases 
with the atomic velocity.
In practice the laser beam is of course always finite, 
so there is a third problem  
aside from reflection and 
delay: 
atomic transmission through the laser without
photon detection, which  mainly affects the fast components and  
may also cause   
%information loss
distortion in the measured photon distribution. 
We shall characterize this effect, and describe several operation  
regimes. Finally, we shall see how, taking advantage of
the Rabi oscillation 
induced by the laser, the flux may also be obtained by
deconvolution with a finite width laser at least in two cases:     
for {\it strong driving} and $\gamma\to 0$, i.e.,
in a limit which is quite the opposite from the one used  
for the semi-infinite laser, and when the momentum width of the atomic 
wave packet is small
compared to its average momentum.

\section{The model}

The setting of the modelled experiment and the fundamental theory are 
described in Ref. \cite{DEHM02} so we shall only outline the basic ideas 
and equations here, emphasizing the novelties   
due to the finite laser-beam width.
A two-level atom wave packet impinges on a perpendicular 
laser beam at resonance with the atomic transition. In the so called
quantum jump approach  \cite{Hegerfeldt93} the continuous
measurement of the fluorescence 
photons is simulated by a periodic projection onto no-photon or 
one-photon subspaces every $\delta t$, a time interval large enough
to avoid the Zeno effect, but smaller than any other characteristic time.
The amplitude for the
undetected atoms in the interaction picture for the internal Hamiltonian 
obeys, 
in a time scale coarser than $\delta t$, and using the
rotating wave and dipole 
approximations, an effective Schr\"odinger equation
governed by the complex ``conditional''  Hamiltonian
(the hat is used to distinguish momentum and position operators from 
the corresponding c-numbers)
\begin{equation}\label{2.7}
H_{\rm c} = \hat{p}^2/2m +\frac{\hbar}{2}\left({0\atop 0}{0 \atop
    -i\gamma} \right) + \frac{\hbar}{2}\,\chi(\hat{x})
\left({0\atop \Omega}{\Omega  \atop 0} \right),   
\end{equation}
where the 
ground state $|1\rangle$ is in vector-component notation ${1 \choose 0}$,
the excited state $|2\rangle$ is ${0 \choose 1}$, 
\beq
\chi(x)=\Bigg\{{ \,1,\;\;\;\; 0<x<L\atop0,\;\;\;\; {\rm otherwise}}
\eeq
and 
$\Omega$ is the Rabi frequency. 
The probability, $N_t$, of no photon detection 
from $t_0$, the instant when the packet is prepared far from the laser 
and with positive momenta, up to time $t$,
is given by \cite{Hegerfeldt93}
\begin{equation} \label{2.5}
N_t = || e^{-i H_{\rm c}(t-t_0)/\hbar} |\psi (t_0)\rangle||^2,
\end{equation}
and the probability density, $\Pi (t)$, for the first photon detection
by
\begin{equation} \label{2.6}
\Pi (t) = - \frac{dN_t}{dt}=\gamma P_2,
\end{equation}
where $P_2$ is the population of the excited state. 
$\Pi$ plays the role of an ``operational'' time-of-arrival distribution. 
%it is the central measured quantity from which we should   
%try to obtain information on the atom. 
To obtain the time development under $H_{\rm
c}$ of a wave packet incident from the left 
we first solve the stationary equation
\begin{equation}\label{eigenvalue}
H_{\rm  c}{\bf \Phi} = E {\bf \Phi},~~~~~{\rm where}~~{\bf
  \Phi}(x)\equiv{\phi^{(1)}(x)\choose\phi^{(2)}(x)} 
\end{equation}
for scattering states 
with real energy 
$
E = \hbar^2k^2/2m \equiv E_k,  
$
which are incident from the left ($k>0$), 
\begin{equation}\label{A15}
{\bf \Phi}_k (x)= \frac{1}{\sqrt{2\pi}}
\left\{\begin{array}{ll}
\left(  
{e^{ikx}+ R_1e^{-ikx} \atop R_2 e^{-iqx}}
 \right), &\quad x\le0, 
\\ 
\left(  
{T_1 e^{ikx}\atop T_2 e^{iqx}}
 \right),& \quad x\ge L. 
\end{array}
\right. 
\end{equation}
These states are not orthogonal, in spite of the reality of $E$, 
because the Hamiltonian $H_{\rm c}$ 
is not Hermitian. The wavenumber $q$ obeys  
\begin{equation} \label{Eq}
E + i\hbar\gamma/2 = \hbar^2q^2/2m,  
\end{equation}
with ${\rm Im}\,q > 0$ for boundedness, while $R_{1,2}$ and $T_{1,2}$
are
reflection and transmission amplitudes yet
to be determined for the ground and excited state  
channels. 

To obtain the form of ${\bf \Phi}_k (x)$ for $0<x<L$ we denote by 
$|\lambda_+\rangle$ and 
$|\lambda_-\rangle$ the (unnormalized and nonorthogonal)
eigenstates of the
matrix $\frac{1}{2} \left( {0\atop \Omega}{\Omega\atop -i\gamma}
\right)$
corresponding to the eigenvalues $\lambda_\pm$, 
\begin{eqnarray}\label{A13}
\lambda_\pm &=& - \frac{i}{4}\gamma \pm \frac{i}{4}
\sqrt{\gamma^2 - 4 \Omega^2}
\\ \label{A14}
|\lambda_\pm \rangle &=&  {1 \choose 2 \lambda_\pm/\Omega}~. 
\end{eqnarray}
(Formally we assume that $\lambda_+\ne \lambda_-$. The case $\lambda_+ =
\lambda_-$ may be treated by taking the limit.) 
For $0<x <L$, one can write ${\bf \Phi}_k$ as a superposition
of $|\lambda_\pm \rangle$ in the form
\beqa
\label{A17}
\sqrt{2\pi} {\bf \Phi}_k (x) &=& C_{++}|\lambda_+ \rangle e^{ik_+x} +
C_{-+} |\lambda_- \rangle e^{ik_-x}
\\
&+&C_{+-}|\lambda_+ \rangle e^{-ik_+x} +
C_{--} |\lambda_- \rangle e^{-ik_-x}
, \qquad 0<x<L, 
\eeqa
which, at variance with the semi-infinite laser case must contain
now growing exponentials in addition to decaying ones. 
From the eigenvalue equation $H_{\rm  c}{\bf \Phi}_k = E_k {\bf \Phi}_k$,
together with $E_k = \hbar^2k^2/2m$, there follows 
\begin{equation}\nonumber
k_\pm^2 = k^2 - 2m \lambda_\pm /\hbar = k^2 + i\gamma m/2\hbar \mp
i  \sqrt{\gamma^2 - 4 \Omega^2}m/2\hbar,
\end{equation}
with Im$\, k_\pm > 0$.
The continuity of ${\bf \Phi}_k(x)$ and its derivative 
at $x = 0$ and $x=L$ leads to eight equations with eight unknowns and 
to explicit but rather lengthy expressions for the coefficients 
of the wave function in the laser region, see (\ref{A17}),
and for the transmission and reflection amplitudes. 
We shall not display them here but discuss different  
limits,  
approximations and numerical examples.  

An important simplification occurs for  
kinetic energies above $\Omega\hbar/2$ or,  
equivalently, for de Broglie wavelengths $2\pi/k$ smaller than  
the Rabi-wavelength $2\pi v/\Omega$:  
the stationary wave may be very well approximated 
according to the classical idea that   
the atom moves along an unperturbed classical 
trajectory, without reflection, and 
with its internal level populations oscillating quantally 
%according to Rabi's frequency, 
in the laser-illuminated region.  
Specifically, 
${\bf \Phi(x)}$  in the laser region ($0<x<L$) can be approximated
as the product of the 
internal wave function for the atom at rest, 
assuming a pure ground state at an entrance instant $t=0$,
and making the substitution $t=x/v$,
times the translational plane wave $e^{ikx}/(2\pi)^{1/2}$,  
\beq
{\bf \Phi}(x)=\frac{e^{ikx}e^{-\gamma x/4v}}{(2\pi)^{1/2}}
\left\{{\cos[\frac{x}{2v}\sqrt{\Omega^2-\gamma^2/4}]
+\frac{\gamma}{2\sqrt{\Omega^2-\gamma^2/4}}
\sin[\frac{x}{2v}\sqrt{\Omega^2-\gamma^2/4}]\atop
\frac{-i\Omega}{\sqrt{\Omega^2-\gamma^2/4}}
\sin[\frac{x}{2v}\sqrt{\Omega^2-\gamma^2/4}]}
\right.
\label{cap}
\eeq
whereas to the left of the laser $R_1\approx R_2\approx 0$, and to the right
the transmission amplitudes $T_{1,2}$
are obtained by continuity from  (\ref{A15}) and (\ref{cap}).
In particular, 
\beq
T_2=e^{i(k-q)L}e^{-\gamma L/4v}
\frac{-i\Omega}{(\Omega^2-\gamma^2/4)^{1/2}}
\sin\left[\frac{L}{2v}(\Omega^2-\gamma^2/4)^{1/2}\right].
\label{te2}
\eeq
The
validity of this semiclassical 
approximation for the translational part of the stationary waves
does not mean that every wave packet with energy components 
well above $\Omega\hbar$ behaves classically. Two or more different 
stationary components may add up coherently in the time dependent 
wavefunction, leading to very non-classical  
interference effects, as we shall see in an
example below.

If
$\widetilde{\psi}(k)$ denotes the wavenumber amplitude that the wave packet
would have as a freely moving packet at $t=0$, then
\begin{equation}\label{2.9}
{\bf \Psi}(x,t) = \int_0^\infty dk \,\widetilde{\psi}(k) \,{\bf \Phi}_k
(x)\,e^{-i \hbar k^2 t/2m}
\end{equation}
describes the ``conditional''  
time development of the state for an undetected atom which in the
remote past comes in
from the left in the ground state.

\section{Operation regimes}

\begin{figure}
%{\includegraphics[width=3.35in,height=2.25in]{ba1.eps}}
{\includegraphics[width=13cm]{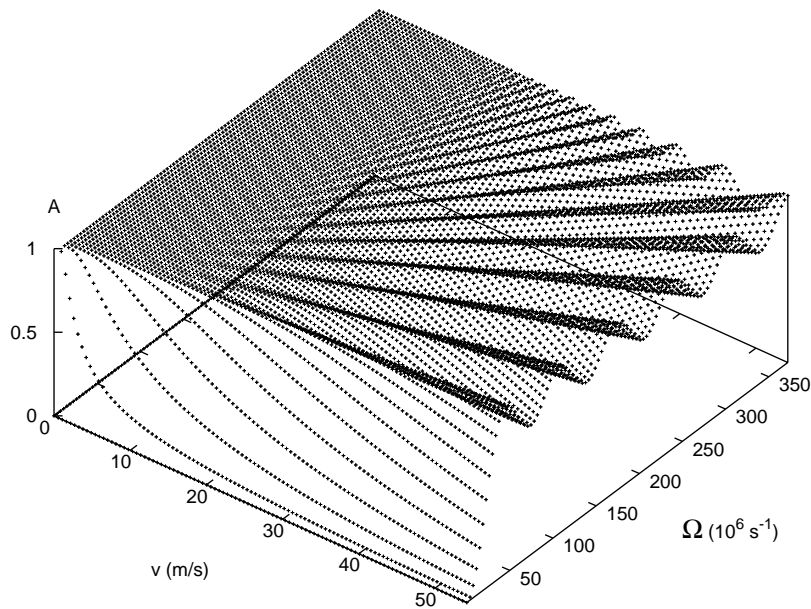}}
\caption{Absorption versus velocity $v$ and Rabi frequency $\Omega$
for $L=5\mu$m.  
This and other figures are obtained for the 
transition at 852 nm of Cs atoms, with
$\gamma=33.3\times 10^{6}$ s$^{-1}$. 
}
\label{abanico}
\end{figure}

\begin{figure}
%{\includegraphics[width=3.35in,height=2.25in]{ba2.eps}}
{\includegraphics[width=13cm]{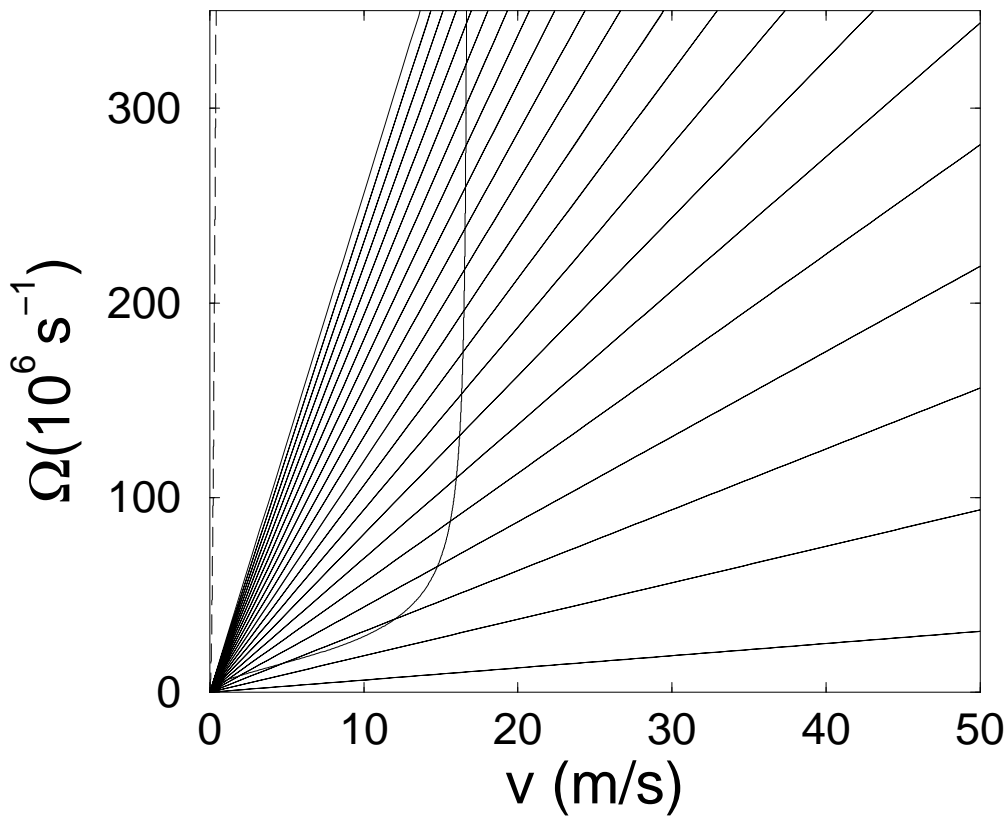}}
\caption{$\Omega$--$v$ plane with separations of the different
regions, $L/l=1$ (curved solid line), $2E/\hbar\Omega=1$ (dashed line), 
and maximum absorption lines up to $n=20$ (straight solid lines), see (\ref{Op}).  
$L=5\mu$m as in figure \ref{abanico}.}
\label{plano}
\end{figure}

\begin{figure}
%{\includegraphics[width=3.35in,height=2.25in]{ba3.eps}}
{\includegraphics[width=13cm]{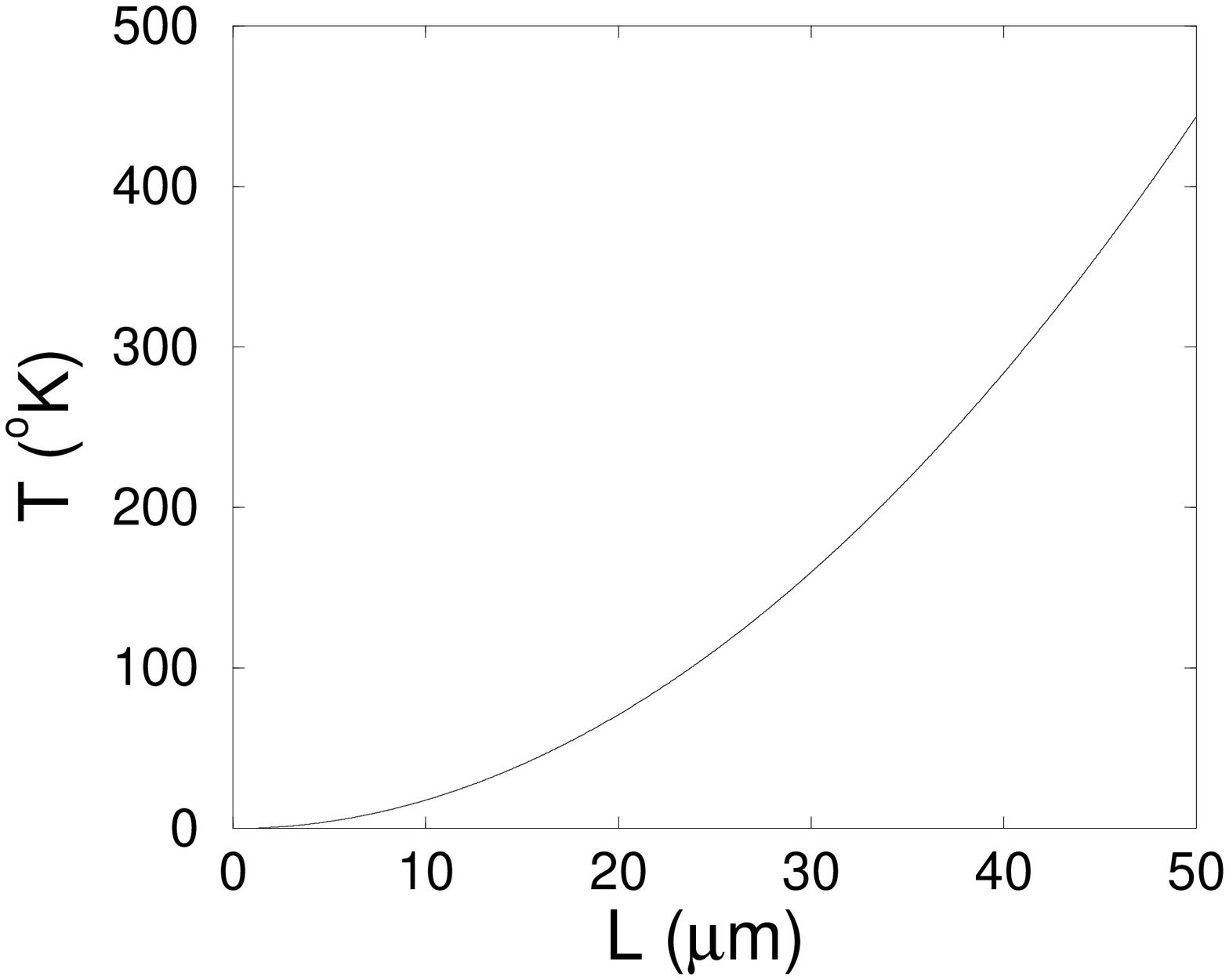}}
\caption{``Critical Temperature'', calculated as $mv_L^2/K$, versus 
laser-beam width $L$ for Cesium atoms and strong driving conditions. 
$v_L=L\gamma/10$, i.e., the velocity for which the penetration length 
is equal to the laser beam width, $l=L$. 
$\gamma=33.3\times 10^{6}$ s$^{-1}$, and $K$ is Boltzmann's constant. 
Above this temperature the effect of the 
finite width of the laser beam cannot be ignored.   
}
\label{lt}
\end{figure}

Figure \ref{abanico} shows the total detection probability, 
or ``absorption'' due to the non-hermitian potential, 
$A=1-|T_1|^2-|R_1|^2$, versus the incident velocity $v$ 
and Rabi frequency $\Omega$ for the stationary
wavevectors ${\bf \Phi}_k$. 
$A$ may also be understood as the probability that the particle be
reflected or transmitted 
in state $ |2\rangle$,
or be detected through a photon emission in the laser region. 
We may distinguish several regions in the $v$--$\Omega$ plane
(these regions are also portraid in figure \ref{plano})
depending on three basic criteria,
each associated with a dimensionless parameter:
\begin{itemize}
\item 
Reflection: Except for very weak driving, where it is not significant, 
reflection is important for kinetic energies below or around $E=\Omega\hbar/2$, 
(dashed line in figure \ref{plano}), 
but  
vanishes when $E>>\Omega\hbar/2$. 
%In absence of reflection, see below, 
%$\Phi_k$ may be approximated semiclassically.   
%In this ``high-energy'' limit  
%a semiclassical approximation is possible, see below. 
\item Laser intensity: Strong driving for $\Omega/\gamma>1$,
and weak driving for $\Omega/\gamma<1$. 
\item Laser beam width: Let $l$ be the
``penetration length'' of the stationary wave in the semi-infinite
laser region, that we may estimate as five times the detection delay 
\cite{DEHM02}
multiplied by the atomic velocity $v$,  
\beq\label{l}
l=5v\left(\frac{2}{\gamma}+\frac{\gamma}{\Omega^2}\right).
\eeq
For $L>l$ the laser behaves effectively as 
a semi-infinite one. This regime,  
to the left of the curved solid line of figure \ref{plano},
corresponds essentially 
to the one examined in the
previous paper. The exact expressions for wave functions, reflection
and transmission amplitudes obtained in \cite{DEHM02} are good approximations
for the finite width laser if $L>l$ but fail otherwise. Figure 
\ref{lt} shows the critical ``temperature'' versus $L$ 
(for Cs atoms and $\gamma=33.3\times 10^{6}$ s$^{-1}$)
above which the finite width of the laser has to be taken into
account. 
\end{itemize}

\begin{figure}
%{\includegraphics[width=3.35in,height=2.25in]{detalle.eps}}
{\includegraphics[width=13cm]{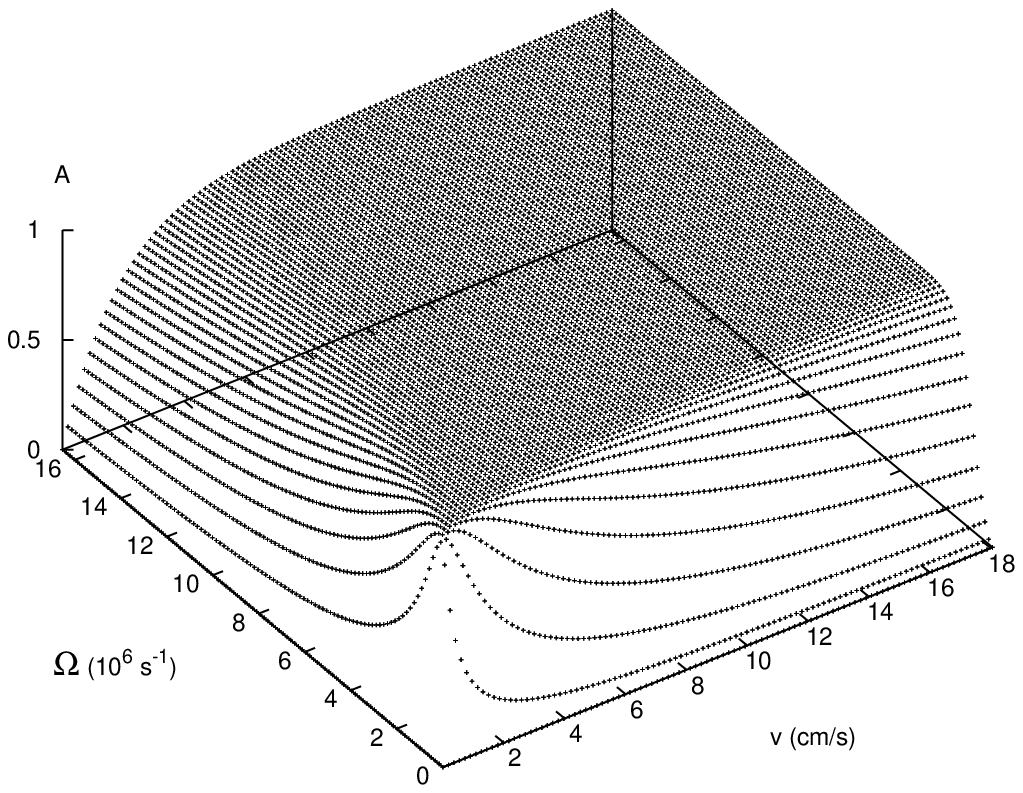}}
\caption{Close-up of figure \ref{abanico}.  
}
\label{detalle}
\end{figure}

For a constant $\Omega$, at moderate to strong driving, $A$ increases 
very sharply from zero to one at low $v$. This increase is not 
visible in figure \ref{abanico} because of the scale, but can be seen 
in the close-up of figure \ref{detalle}.
The small absorption near $v=0$ is due
to reflection into the ground state.
After the rapid increase
of $A$ at low $v$ along a constant $\Omega$ line, there follows a
{\it plateau} of perfect 
detection, still within the ``semi-infinite laser'' regime,
see again figure \ref{abanico}. 
Gradually, 
for higher and higher velocity $v$ and in  
strong driving conditions,
the plateau is substituted by an oscillating pattern. 
The absorption surface  
to the right of the plateau resembles a fan,  
which can be explained in terms of the  
semiclassical picture 
and the finiteness of the laser ``barrier''. The ridges
correspond to 
combinations of parameters where the Rabi oscillation leaves  
the excited state population 
at a maximum on the laser edge $L$, so that the decay takes place behind 
the laser,  
at $x>L$, with a detection delay $1/\gamma$ from the instant when the 
classical trajectory leaves the laser.    

\begin{figure}
%{\includegraphics[width=3.35in,height=2.25in]{bien.eps}}
{\includegraphics[width=13cm]{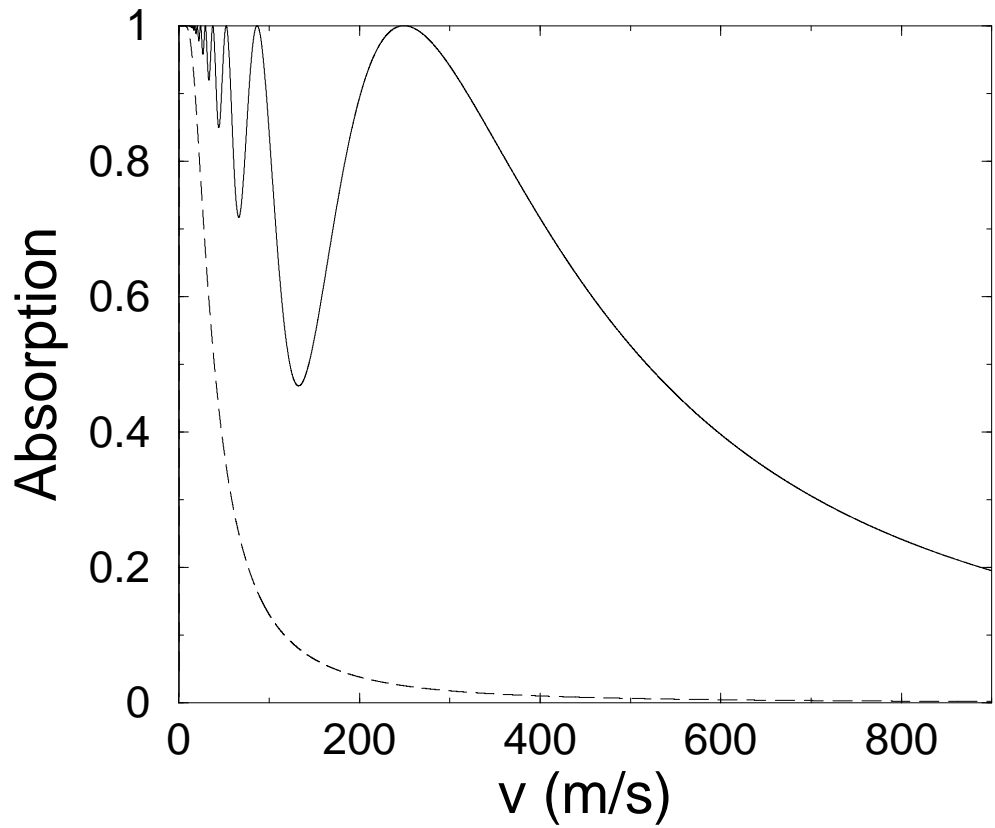}}
\caption{Absorption versus $v$ for $\Omega=5 \gamma$ (solid line), 
and $\Omega=\gamma/2$. $L=5 \mu$m. 
}
\label{bien}
\end{figure}

Each ridge corresponds to a number $n+1/2$ of Rabi oscillations
between $x=0$ and $x=L$, and will be indexed by $n=0,1,2...$. 
Neglecting the effect 
of $\gamma$ within the laser region, the velocity at the ridge for a given
$\Omega$ is 
given by 
\beq\label{on}
v_n=\frac{L\Omega}{(2n+1)\pi}.
\eeq
Equivalently, each absorption ridge may be characterized by the
straight line 
\beq\label{Op}
\Omega_n=\frac{(2n+1)\pi v}{L}
\eeq
in the $\Omega$--$v$ plane.  
The first one, for $n=0$,  
corresponds to a spatial version of a $\pi$ pulse, i.e., the velocity, 
laser-beam width, and Rabi frequency are just the right ones to pump the 
(semiclassical) atom, 
from the ground state at $x=0$ to the excited state at $x=L$,
in half Rabi 
oscillation.   
Note from  (\ref{on}) that the fan structure ``folds upwards''
by decreasing $L$, 
because smaller lengths require higher laser intensities to achieve 
full detection.   
The minima at the valleys correspond to an entire number of Rabi 
oscillations. These minima are not exactly zero due to
the absorption
that takes place 
in the laser region. As the velocity $v$ increases,
however, this absorption
decreases, as may be seen in figure \ref{bien}.
Finally, for velocities greater than $v_0$, the atom is too fast for 
being completely excited during the crossing of the laser-illuminated 
region, and the absorption decreases monotonously. The solid line 
in figure \ref{bien} shows a cut of the absorption surface 
along a constant-$\Omega$ line for strong driving conditions.

The velocity window 
for nearly complete detection on a ridge along a 
constant-$\Omega$ line  
is larger for smaller $n$. This is also clearly illustrated in figure 
\ref{bien}. 
A simple formula for the window width 
$\Delta_W$ around the maximum 
where absorption is better than 
$99\%$ is   
\beq
\Delta_W <\left[\frac{2}{\pi(2n+1)}\right]^2\frac{L\Omega}{10}
=\frac{4 v_n}{10 \pi (2n+1)}. 
\label{pwidth}
\eeq  
Notice that even in the best case, $n=0$, a wave packet located 
essentially 
within one of these windows of full detection will have
a small width $\Delta v$ 
compared to its average velocity. In particular, for $\Delta_W=8\Delta v$,
\beq
\label{window}
\Delta v\approx \frac{v_n}{(2n+1)60}. 
\eeq
An interesting feature of the transmission amplitude 
$T_2$ at the detection maxima is its alternating sign, 
$-i$ for $n$ even, and $i$ for $n$ odd, see (\ref{te2}).
We shall comment on  physical
consequence below. 
 
The above description is not applicable to weak driving
conditions.   
Reflection in particular is less important at low velocities
than for strong driving, and it 
vanishes for any $v\ne 0$ when  $\gamma/\Omega\to \infty$.   
Also, the plateau of full detection is much narrower than for
strong driving,  
and for higher velocity $v$ along a 
constant-$\Omega$ line, 
the laser does not have enough 
intensity to produce Rabi oscillations, 
so the 
absorption decays monotonously with $v$, see figures \ref{abanico},
\ref{detalle}, and \ref{bien}.  

%
%\section{Deconvolution for strong driving conditions} 
%
The detection delay for a given incident wavenumber will in general 
have contributions from the laser region and from the region 
outside. The delays in the laser region are $2/\gamma$ for strong 
driving or $\gamma/\Omega^2$ for weak driving when $L>l$ \cite{DEHM02}, 
or fractions of these quantities for smaller laser-beam widths, whereas  
the detection delay once the (semiclassical) excited atom arrives  
at $L$ is $1/\gamma$. This later contribution 
would be very small for large (ideally infinite) $\gamma$,
but in this limit, with all other parameters constant,   
there would not be excitation at all.  
One might then imagine the case where both $\gamma$ {\it and} $\Omega$
are infinite,  
but with a constant ratio $\Omega/\gamma$.
In this case, however, 
the penetration length goes to 
zero, and the semi-infinite laser result, $R_1 \to -1$, applies, 
so that all atoms would be reflected without
being excited \cite{DEHM02}.  
Another not very helpful limit is $L\to 0$, with all other parameters 
fixed, because all atoms are transmitted in the ground state  
without being detected\footnote{One might of course insist on using the 
information of the very few detected atoms and normalize,
but the result is  strongly biased in favour of 
the slowest energy components.}. 
``Good'' direct (no delay, full detection) measurements are possible, 
but only within specific parameter 
ranges.  
For moderate velocities, i.e., provided $L>l$ for the   
wave packet momentum components, and above the 
reflection region, the measurement 
on the full-detection plateau can be very efficient, 
as illustrated in Ref. \cite{DEHM02}.
This requires, for strong or weak driving,
\beq
\frac{\hbar}{E}\ll\frac{1}{\Omega}+\frac{\gamma}{\Omega^2}\ll \frac{2}{\gamma}
\ll\frac{L}{5v},\Delta t,  
\eeq
where $\Delta t$ is the time span of the wave packet passage. 
%Similarly, for weak driving, 
%
%\beq
%\frac{\hbar}{E}\ll\frac{\gamma}{\Omega^2}\ll \frac{2}{\gamma}
%\ll\frac{L}{5v},\Delta t. 
%\eeq
%
Even when these conditions are fulfilled, the observed $\Pi$ 
is only approximately equal to $J$.  
Unfortunately, 
the deconvolution used in Ref. \cite{DEHM02} 
to get the flux $J$ exactly for large $\gamma$ may be
impossible to
implement with a finite-width laser at large 
velocities, since the penetration length $l\to\infty$ as
$\gamma\to\infty$. The following section provides a way out   
which, again, is not universal but depends in fact on  
rather restrictive 
conditions.

\section{An ideal distribution}
Full detection is achieved sufficiently close to the ridges of the fan 
structure for moderate to strong driving.  
The first maximum, corresponding to $n=0$,
is particularly suitable because it 
provides the largest 
momentum window of nearly 
complete absorption. 
Let us assume that   
$\Omega$ and $L$ may be adjusted so that most of a wave packet    
lies within an absorption maximum.  
As in the weak driving case of the semi-infinite laser,
full detection comes
with a price: the detection delay due to the time necessary 
to de-excite the atom, now  outside the laser illuminated region.    
We have pointed out already that a short lifetime (a large $\gamma$)   
``solves'' this problem but also creates a new one: the absence 
of excitation and therefore of detection. This suggests that    
the useful limit may be quite the opposite, 
namely, $\gamma\to 0$. The long delay can then be substracted 
using a convolution formula to obtain an ``ideal'' distribution that 
coincides with the flux when $\gamma\to 0$, or, quite independently 
of $\gamma$, 
for 
narrow wave packets in momentum space.

To define from the  
``experimental'' $\Pi(t)$ an idealized arrival-time distribution,
we shall assume 
\beqa\label{ineq}
\frac{\hbar}{E}
<<\frac{1}{\Omega}{\stackrel{<}{\sim}}\frac{L}{v}<<\frac{1}{\gamma},
\eeqa
so that we can use the semiclassical and
strong driving approximations neglecting
the effect 
of $\gamma$ between 0 and $L$, since the only significant absorption 
(i.e., detection) occurs for 
$x>L$.

$\Pi_{\rm id}(t)$ is then defined by the convolution formula
\begin{equation} \label{4.1}
\Pi = \Pi_{\rm id}*W, 
\end{equation}
where  
\begin{equation}\label{A4}
W (t) = \gamma e^{-\gamma t} 
\end{equation} 
is the probability density to detect a photon at time $t$ 
if the atom is excited at $t=0$ in 
the laser-free region.  
Even though (\ref{4.1}) is in fact the equation that {\it defines}
$\Pi_{\rm{id}}$, there is 
of course a ``classical'' physical motivation for the convolution
structure.
If the arrival time 
of the atom,
and the time of photon emission from the arrival instant were independent
random variables, then the distribution for the sum of these two quantities 
would have precisely that form. It turns out, as discussed in Ref. \cite{DEHM02} that
$\Pi_{\rm{id}}$ may become negative so, 
generically,
this simple hypothesis does not hold true quantum mechanically. This  
does not invalidate $\Pi_{\rm{id}}$, but should make us cautious about its
interpretation: it {\it plays the role} of a
time of arrival density in quantum mechanics, but does not share all 
the properties of the corresponding classical density, 
most prominently positivity. 
This resembles the status of the Wigner function, which plays the role 
of a classical phase-space probability density but can also
be negative.

\begin{figure}
%{\includegraphics[width=3.35in,height=2.25in]{jfk2.eps}}
{\includegraphics[width=13cm]{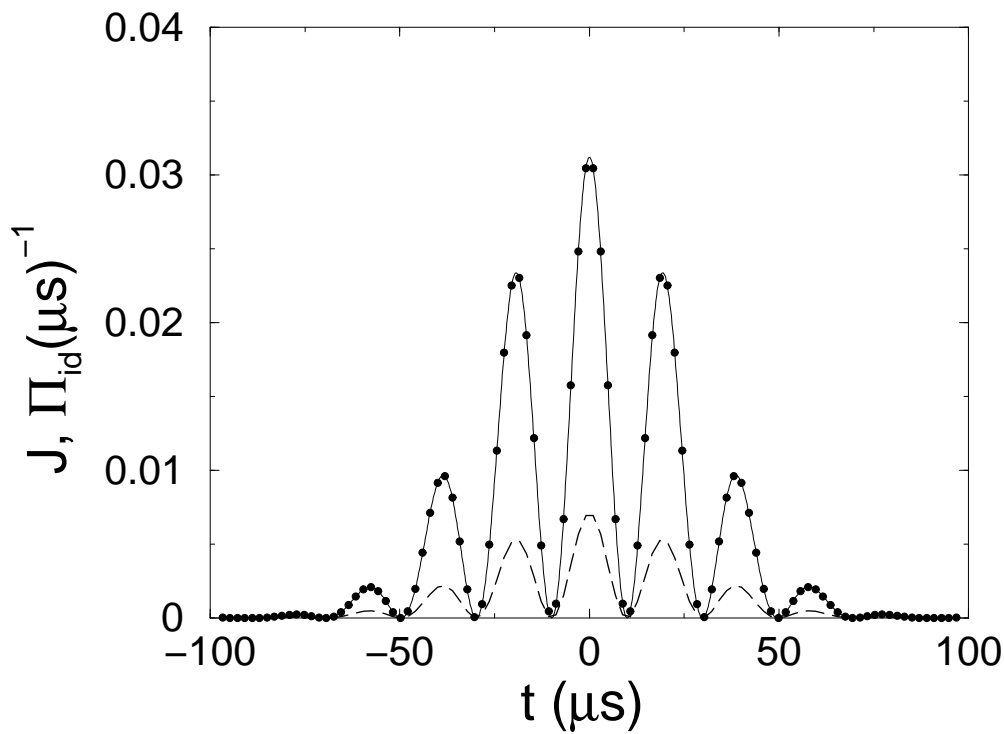}}
\caption{Flux (solid line), $\Pi_{\rm id}$ (dashed line), 
and normalized $\Pi_{\rm id}$ (dots), 
for the coherent sum of two Gaussians with average velocities 
$v_1=167.05$m/s and $v_2=v_1+.9\mu$m/s, and $\Delta x= 4233 \mu$m. 
They become minimal-uncertainty-product states when their centers
arrive at $L=5\mu$m. $\Omega=104.43\times 10^6$ s$^{-1}$. 
}
\label{jfk2}
\end{figure}

The ideal distribution $\Pi_{\rm id}$ is obtained  
from the relation between Fourier transforms, 
$\widetilde{\Pi}_{\rm id} =
\widetilde{\Pi}/\widetilde{W}$ where ${\widetilde \Pi}(\nu) = \int dt
e^{-i\nu t} \Pi (t)$ etc. From (\ref{A4}) one finds
\begin{equation}
\label{A5}
\widetilde{W} (\nu) = \frac{\gamma}{ 
i \nu +\gamma},   
\end{equation} 
so that, in the 
time domain, 
\begin{equation}\label{4.1a}
\Pi_{\rm id}(t) = \Pi (t) + \frac{1}{\gamma} \Pi'(t). 
\end{equation}
%
%Adjusting $L$ and $\Omega$ the wave packet central momentum 
%may be centered at one of the absorption maxima $p_n$, preferably 
%$p_0$.
If the wave packet is centered at one of the absorption maxima, 
and the momentum width satisfies (\ref{pwidth}), the atoms are all 
pumped
to the excited state,  
$|T_2|\approx 1$,  and we may write 
\begin{equation}\label{A3}
{\Pi}(t)
= \frac{\gamma}{2\pi} \int_L^\infty dx  \int dk\int dk'\, {{\widetilde
\psi} (k)} 
{\widetilde\psi}^* (k')  
e^{-i(E-E')t/\hbar}e^{i(q-q'^*)x}. 
\end{equation}
Performing the $x$ integral, and approximating
the resulting denominator according to
(\ref{ineq}), 
%in the resulting denominator
%(in the exponent the first term is enough because 
%of Eq. (\ref{ineq})),  
%
\beq
q-{q'}^*\approx k-k' +\frac{i\gamma m}{2\hbar}\frac{k+k'}{kk'}, 
\eeq
one finally gets 
\beqa
\Pi_{\rm id}(t)&=&\frac{1}{2\pi}\int\int dk dk' 
e^{-i(E-E')t/\hbar} e^{i(k-k')L} \widetilde{
\psi} (k) 
\widetilde{\psi}^* (k')  
\nonumber
\\
&\times&
\bigg[
%e^{-\frac{\gamma m L}{2\hbar}\frac{k+k'}{kk'}}
\frac{i\gamma+(k^2-k'^2)\hbar/2m}{(k-k')+\frac{i\gamma m}{2\hbar}
\left(\frac{k+k'}{kk'}\right)}\bigg]. 
\eeqa
The term in square brackets tends to the kernel of the 
integral that gives the current density $J$, 
\beq\label{kj}
\frac{(k+k')\hbar}{2m},
\eeq
in several limits. Using $k'=k+\delta$, this occurs in particular 
for $\gamma\to 0$, and also for $\delta/k\to 0$ independently
of the value of $\gamma$, as long as (\ref{ineq}) is satisfied.    
In this later case, namely for a small momentum width in comparison to the 
momentum itself, (\ref{kj}) coincides also with the kernel  
of Kijowski's axiomatic TOA distribution, 
\beq\label{ki}
\frac{(kk')^{1/2}\hbar}{m},  
\eeq
since $k\approx k'$ and the arithmetic and geometric means 
in  (\ref{kj}) and (\ref{ki}) coincide. 
The condition of small momentum-width also implies that
the effect of the derivative in (\ref{4.1a}) is negligible, and that, 
even if the last inequality of (\ref{ineq}) fails,  
$J$ can be obtained very accurately by normalizing the observed results.  
An example is shown in figure 
\ref{jfk2}, where two Gaussians on the top of a single ridge 
have been added coherently,
\beq
\label{ab}
\psi=\psi_a+\psi_b,
\eeq
to produce 
an interference pattern in the flux.

%\subsection{Observation of backflow}
%
According to (\ref{window}), the condition $\Delta v/\la v\ra<<1$
is necessarily fulfilled  
when the wave packet momenta are within one ridge of maximum 
detection,
so that 
the flux and 
Kijowski's distribution will coincide, i.e., no negative values of   
$J$ may be found in this case. In order
to measure 
backflow the wave packet must be located within the full detection 
plateau and then proceed as in Ref. \cite{DEHM02} but  
the finite width of the laser beam  
imposes a maximum speed for this method to work.   
Another route would be to use different 
ridges of maximum detection for each gaussian. The two gaussians  
should be on odd or even $n$ ridges but not on contiguous 
ones  because of the sign change of the transmission amplitude
mentioned below (\ref{window}).
%The problem though is that 
%the energy differences would be high that the time scale of 
%the interference would oscillate too rapidly for a simple 
%experimental test. 

\begin{figure}
%{\includegraphics[width=3.35in,height=2.25in]{jfk2.eps}}
{\includegraphics[width=13cm]{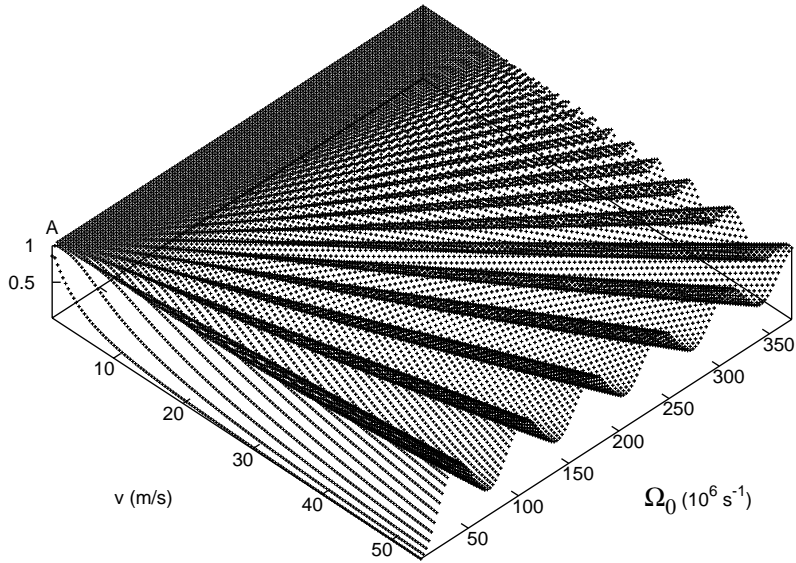}}
\caption{
Absorption versus velocity $v$ and $\Omega_0$
for $L=5\mu$m
and position dependent Rabi frequency, 
$\Omega(x)=L\Omega_0 e^{\frac{-(x-x_0)^2}{2\delta^2}}/\delta (2\pi)^{1/2}$; 
$\delta=0.529\mu$m, $x_0=2.5\mu$m, and   
$\gamma=33.3\times 10^{6}$ s$^{-1}$.
The calculation is performed with a transfer matrix method. 
}
\label{abag}
\end{figure}

\section{Conclusions}\label{concl}
In this paper we have analyzed the measurement 
of the arrival time of an atom at a given 
location by means of a laser that illuminates a
finite region.  
Repeating the experiment many times for a given 
atomic preparation produces an ``observed 
distribution'' of first fluorescence photons $\Pi$,
that will in general 
be distorted with respect to the ``axiomatic'' distribution of Kijowski
$\Pi_K$ or the flux $J$ (both evaluated for the isolated atom,
i.e., without the laser) because of 
detection delay, and atomic reflection or 
transmission without photon detection.
(In general even if the three problems are made negligible $\Pi$
is only approximately
equal to these ``ideal'' quantities.) 
%Whereas the distortions caused by 
%defficient detection of lower or higher tails of the wave packet momenta
%are generically fatal and cannot be corrected, since the
%necessary information is lost 
%(the only exception is the case where undetected reflection and or
%transmission 
%are uniform in the relevant momentum width of the packet), 
It is possible to correct for the 
distortions due to detection delays, at least in certain limits, and obtain 
the current density by deconvolution.      
In the present model the laser beam intensity has sharp edges  
to facilitate the analytical examination but very similar 
results have been obtained for a more realistic Gaussian profile, 
see figure \ref{abag}.

\ack{
This work has been supported
by Ministerio de Ciencia y Tecnolog\'\i a (BFM2000-0816-C03-03), 
UPV-EHU (00039.310-13507/2001),
and the Basque Government (PI-1999-28).}

%\newpage

\end{document}